\documentclass[a4paper, amsfonts, amssymb, amsmath, reprint, showkeys, nofootinbib, twocolumn,superscriptaddress]{revtex4-1}
\usepackage[english]{babel}
\usepackage[utf8]{inputenc}

\usepackage{amsthm}
\usepackage{mathtools}
\usepackage{physics}
\usepackage{xcolor}
\usepackage{graphicx}
\usepackage[left=25mm,right=25mm,top=35mm,columnsep=15pt]{geometry} 
\usepackage{adjustbox}
\usepackage{placeins}
\usepackage[T1]{fontenc}
\usepackage{lipsum}
\usepackage{csquotes}

\usepackage[pdftex, pdftitle={Article}, pdfauthor={Author}]{hyperref} 
\begin{document}

\title{Studies of thorium and ytterbium ion trap loading from laser ablation for gravity monitoring with nuclear clocks}

\author{Marcin Piotrowski}
\email{marcin.tutmo@gmail.com}
\address{CSIRO Manufacturing, Pullenvale, Queensland 4069, Australia}
\address{Centre for Quantum Dynamics, Griffith University, Brisbane, QLD 4111 Australia}
\address{Currently at French-German Research Institute (ISL), 68301 Saint-Louis, France}

\author{Jordan Scarabel}
\address{Centre for Quantum Dynamics, Griffith University, Brisbane, QLD 4111 Australia}
\author{Mirko Lobino} 
\address{Centre for Quantum Dynamics, Griffith University, Brisbane, QLD 4111 Australia}
\address{Queensland Micro- and Nanotechnology Centre, Griffith University, Brisbane, QLD 4111, Australia}
\author{Erik Streed} 
\address{Centre for Quantum Dynamics, Griffith University, Brisbane, QLD 4111 Australia}
\address{Institute for Glycomics, Griffith University, Gold Coast, QLD 4222 Australia}
\author{Stephen Gensemer}
\address{CSIRO Manufacturing, Pullenvale, Queensland 4069, Australia}
\address{Centre for Quantum Dynamics, Griffith University, Brisbane, QLD 4111 Australia}




\begin{abstract}
Compact and robust ion traps for thorium are enabling technology for the next generation of atomic clocks based on a low-energy isomeric transition in the thorium-229 nucleus. We aim at a laser ablation loading of single triply ionized thorium  in a radio-frequency electromagnetic linear Paul trap. Detection of ions is based on a modified mass spectrometer and a channeltron with single-ion sensitivity. In this study, we successfully created and detected $^{232}$Th$^+$ and $^{232}$Th$^{2+}$ ions from plasma plumes, studied their yield evolution, and compared the loading to a quadrupole ion trap with Yb. We explore the feasibility of laser ablation loading for future low-cost $^{229}$Th$^{3+}$ trapping.
The thorium ablation yield shows a strong depletion, suggesting that we have ablated oxide layers from the surface and the ions were a result of the plasma plume evolution and collisions. Our results are in good agreement with similar experiments for other elements and their oxides. 
\end{abstract}

\maketitle
\section{Introduction}

Gravity can be monitored with atomic clocks through a technique known as chronometric leveling \cite{vermeer_chronometric_1983}. It is a well-known consequence of general relativity that clocks sitting in different gravitational potentials have different ticking rates. Clock networks can provide a new way for ground-based gravimetry at the Earth's surface, allowing us to monitor minute drifts in the gravity anomalies. In recent years, rapid progress in atomic clock networks makes this kind of gravity sensor feasible\cite{0034-4885-81-6-064401}.  Such networks have been demonstrated with two state-of-the-art optical atomic clocks compared via fibre link \cite{takano_geopotential_2016,lisdat_clock_2016,chou_optical_2010,PhysRevLett.118.221102}, including demonstrations of a transportable clock \cite{Grotti2018}. The best of lab-based optical clocks are capable of detecting down to 1~cm differences between clock altitudes, which is already beyond the geodetic limit \cite{McGrew2018}. The optical fiber links are no longer a limiting factor for such networks, as stability of frequency transfer down to $10^{-20}$ has been demonstrated \cite{Predehl441}. Even optical links hundreds or thousands of kilometers long can perform well enough for requirements of chronometric leveling with millimeter precision \cite{Predehl441,Lopez:12}. The portability of optical clocks and maintaining their performance out-of-the lab is one of the main obstacles we face in the development of a true gravity observing clock network. While clocks with systematic uncertainty below $10^{-18}$ have been developed in a laboratory setting \cite{Brewer_2019}, portable optical clocks have not reached the accuracy and stability of laboratory-based clocks \cite{koller_transportable_2017}, but advances in the field are being made rapidly \cite{takamoto_test_2020}.
 
 The Thorium-229 nucleus possesses a metastable excited state, $^{229\mathrm{m}}$Th, which exhibits the lowest known excitation energy of all nuclei. The excitation of this transition from the ground state requires energy around 8~eV (the corresponding wavelength is ~150~nm) \cite{Beck_energy_2007,Beck_imroved_2009,Seiferle2019,Yamaguchi_2019,sikorsky2020measurement}. The isomer's existence was inferred based on indirect experimental evidence more than forty years ago \cite{KROGER1976}. Its exact energy value was debated over the years before the recent final isomer's direct detection via the internal conversion decay channel \cite{Seiferle2019}. New ways to precisely measure the frequency of the nuclear clock transition are also emerging \cite{Rellergert2010,Porsev2010,VonDerWense2017,Verlinde2019,PhysRevC.100.044306,Masuda2019,Bilous2020,VonderWense2020}, with the electron bridge process having application to $^{229}$Th$^{3+}$ \cite{Porsev2010_2,Bilous2018}. A clock operating on this transition is an approach pursued to overcome the challenges of present optical clocks \cite{Peik_2003}.  Its narrow linewidth is uncoupled from most of the environmental effects that plague existing atomic clocks. A single $^{229}$Th ion housed in an electromagnetic trap is expected to allow frequency measurements of unprecedented accuracy reaching $10^{-19}$\cite{campbell_single-ion_2012}. That is almost an order of magnitude better than the best optical electronic-shell clock in use now, namely the clock based on $^{27}$Al$^+$ with a systematic uncertainty of $9.4\times 10^{-19}$ and frequency stability of $1.2\times 10^{-15}/\sqrt{\tau}$ \cite{Brewer_2019}. On top of that, the single-ion clock has great potential to be highly miniaturized \cite{Delehaye-2018}. The $^{229}$Th$^{3+}$ system is also attractive from several technical perspectives \cite{PEIK2015516}, design and simulation work has shown that a trapped Th$^{3+}$ ion could be developed in a compact and relatively low-cost package, making field deployment in a dispersed network achievable. In addition, the $^{229}$Th nucleus is expected to be an excellent system for the search for time variation of fundamental constants \cite{Flambaum_enhenced_2006,Berengut_Flambaum_2010,Thirolf_2019}. To this date, a few groups worldwide have demonstrated thorium trapping in ion traps \cite{PhysRevLett.102.233004,Campbell2011, 1807.05975, Borisyuk-2017, Borisyuk-2017-2, PhysRevA.88.012512,PhysRevA.85.033402, thielking_laser_2018}.

The nuclear clock is not the only stimulating application of thorium ion traps. Single ions can be used as very sensitive probes of the electric field, as was recently demonstrated for ytterbium ions \cite{Blumseaao4453}. A potential application of such single ion-based force sensors is the manipulation and investigation of dipole moments in bio-molecules \cite{Streed-unfolding}. The UV light used in the ytterbium ion force sensor can be harmful to bio-molecules, and thus might not be very useful. In contrast, Th$^{3+}$ ion has atomic transitions used for cooling and imaging in the red and infrared parts of the electromagnetic spectrum \cite{PhysRevLett.102.233004}, making it very attractive as a potential force sensor for large bio-molecules. 

Laser ablation of solids is a technique widely used in material processing and deposition where precise and effective material removal from the surface is crucial \cite{willmott_pulsed_2000}. This technique can also deliver ions to an RF trap by vaporization and ionization of material from the metal surface \cite{knight_storage_1981}. Short laser pulses in the nanosecond range are the most effective in terms of the level of ionization of the ablated material \cite{olmschenk_laser_2017}. Nanosecond laser ablation offers temporal control over trap loading and is a particularly attractive method for experiments with micro traps \cite{hendricks_all-optical_2007, cao_compact_2017, Vrijsen:19}. The trap loading of ions into RF traps was demonstrated for Ca \cite{sheridan_all-optical_2011}, Sr \cite{leibrandt_laser_2007}, and Th atoms \cite{zimmermann_laser_2012,troyan_generation_2013}. Systematic studies of the ablation process and trap loading are challenging because of changes in the metal surface caused by the removal of the material from the spot. After repeated pulses, a crater in the metal surface forms and the ablation yield diminishes. To describe this process quantitatively we performed a series of measurements where the number of ions from the same spot was monitored over long times. Our findings of the evolution of the yield from the laser ablation are complementary to previous findings, and we present a more thorough examination of pulse yield depletion.

Here, we present efforts to develop effective methods for ion trap experiments with thorium-229 isotope and future studies of nuclear clock transition. One of the main difficulties with the $^{229}$Th isotope is that it is radioactive with a relatively short half-life of fewer than 8000 years compared to the much more stable $^{232}$Th isotope (a half-life of $1.4\times 10^{10}$~years). The latter is practically much easier to handle and obtain, that is why we applied it to examine methods of target preparation and trap loading. We also investigated Yb for tests of experimental procedures, because more experimental and theoretical data is available about this element in terms of ion trapping and laser ablation. We used ytterbium along with thorium for comparison of laser ablation yield evolution and trap loading efficiency for different charge states of both elements.

\section{\label{sec:apparatus}Methods and experimental setup}

\begin{figure*}
    \centering
    \includegraphics[width=0.8\textwidth]{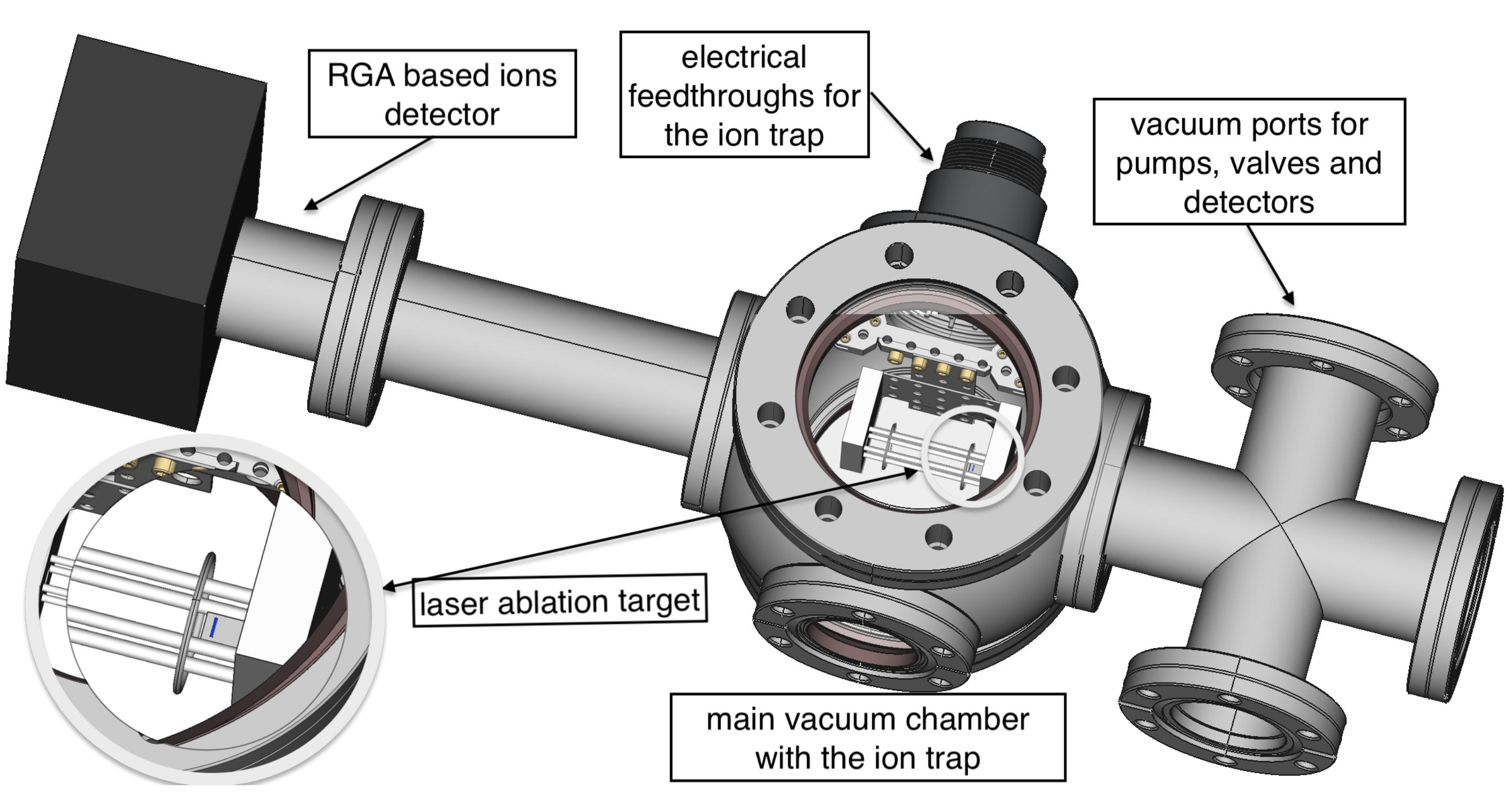}
    \caption{A rendering of the experimental setup. The ion trap is placed inside the central vacuum chamber. The port on the left of the chamber leads to the RGA based ion detection system. The axis of the ion trap is aligned with the axis of the RGA's mass filter rods. The port on the right-hand side connects to the vacuum control and detection system. Top and bottom 4.5-inch ports and front smaller port are closed with windows for optical access. The rear smaller port serves for electrical connection to the ion trap.}
    \label{fig:ion-trap-vacuum}
\end{figure*}

The experimental set-up consists of an ion trap with an ion detector and a laser system for ablation. The ion trap, the ablation targets, and the ion detection system are under ultra-high vacuum, pumped by turbomolecular and ion pumps. 

Fig. \ref{fig:ion-trap-vacuum} shows the rendering of the experimental setup. A spherical cube vacuum chamber (Kimball Physics) holds the ion trap and laser ablation targets. The parts are mounted by grove grabbers to the internal grooves of the vacuum ports. The optical access to the ion trap is provided through the top and bottom $4.5$~inch diameter ports and one smaller side port. The viewports are equipped with anti-reflection coated glass windows. The vacuum pumps, gauges, and detection system are connected by three 2.75 inch ports. The remaining fourth port allows additional optical access for laser cooling and probing of ions in the trap.

\subsection{Ion trap}

Our ion trap is a standard radio-frequency (RF) linear Paul trap \cite{RevModPhys.62.531} modified from the one formerly used for experiments with Yb ions \cite{Kielpinski:06,Streed-2008}. The rod diameters, spacing, and other crucial dimensions are listed in table \ref{paulprop}. The quadrupole radius refers to the distance from a geometric center of a square formed by the quadrupoles (trap axis) to a quadrupole rod. The quadrupole length is the length of the rods. The wire radius is half the wire gauge used for the quadrupole and end caps. The end cap distance from the center is half the distance between the two end caps. The end cap radius is the distance from the perimeter of the circular end cap nodes to their center. The RF field is provided from a digital signal generator (RIGOL DG4162), amplified by 30 dB, and transmitted to the trap via a helical resonator for impedance matching. The trap and helical resonator are resonant at a frequency around 6.2 MHz, allowing us to confine different ionic states of thorium and ytterbium by tuning the amplitude of the drive RF signal and the DC bias applied between adjacent rods.

\begin{table}
\begin{center} 
\begin{tabular}{ | l | l | l | }
Quadrupole rods radii & $r_0$ & 4.508 mm \\ 
Quadrupole rods length & $l_0$ & 51.5 mm \\ 
Rods and end cap wire radius & $r_w$ & 787.5 $\mu$m \\ 
End cap distance from centre & $z_0$ & 20 mm \\
End cap radius & $r_z$ & 7 mm
\end{tabular} 
\caption{Dimensions of the Paul trap.}
\label{paulprop}
\end{center}
\end{table}

\subsection{Laser system for ablation of metal targets} 

The laser ablation as a method for trap loading is most commonly applied to elements like thorium with very high  evaporation temperature. Our laser plasma ablation setup is similar to the one described in \cite{zimmermann_laser_2012}. 
The ablating laser source is a nanosecond pulsed nitrogen laser (Stanford Research Systems NL100). It produces a 3.5 ns pulse with an energy of 170 $\mu$J. The wavelength is 337.1~nm. The repetition rate of the laser can be tuned between 1~-~20~Hz. The output beam was focused to a spot of 350~ $\mu$m $1/e^2$ diameter. The laser beam enters via the top window of the chamber and is focused on the targets that are to be ablated. Taking into account the absorption coefficient of thorium, our pulsed laser generates a maximum fluence of about 0.35~J/cm$^2$. 

\subsection{Thorium and ytterbium targets}

Thorium-229 isotope metal samples are necessary for future nuclear clocks, that is why we carried out a series of preliminary experiments to establish a simple and reliable method of target preparation. The thorium electrodeposition on a stainless steel plate was adopted from the procedure used for uranium in \cite{Dumitru2013}. We used a 0.02M thorium nitrate water solution to electrodeposit thorium on a stainless plate (Kimpball Physics eV parts). A deposition set-up was composed of a platinum electrode, the ammonia hydroxide was used for pH control. According to weighing the plate before and after the process we deposited 0.05 g of material on the plate. A laser ablation signal from a target prepared in this way was observed visually and used for tests of our detection system at that time. A thorium signature was observed on a mass spectrometer after the ablation of the target, so this simple method of target preparation may serve for $^{229}$Th ion loading in the future. Even though targets of thorium-232 prepared this way were not used eventually for the results presented here, the findings are of general importance for future compact thorium-229 ion traps.

For all the results presented here, we changed to pure thorium-232 metal.  A piece of thorium wire approximately 1 mm long was spot-welded to the surface of an aluminum strut inserted throughout the ceramic holder. The thorium 0.22~mm wire is naturally monoisotopic thorium with 99.5\% purity (The Goodfellow supplier). Similarly, a piece of ytterbium wire with natural abundance is placed by the side of the Th target. 

\subsection{Ion detection and plasma diagnostics}

\begin{figure}
    \centering
    \includegraphics[width=\linewidth]{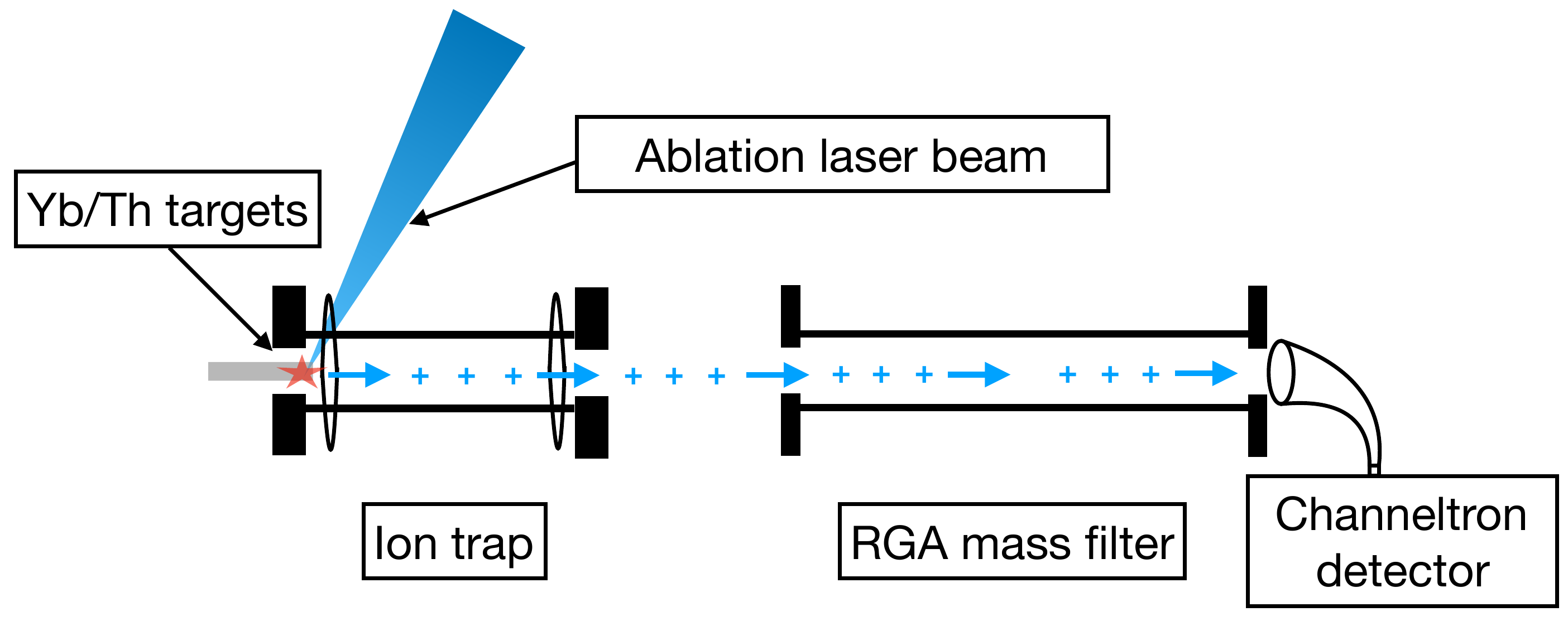}
    \caption{A schematic illustrating the relative positions of the laser plasma creation, the ion trap, and RGA based ions detector with the channeltron. The ions created in the ablation process travel along the path marked towards the detector.}
    \label{fig:RGA_ion}
\end{figure}

We developed a simple non-optical detection system that was sensitive enough to detect a single ion signal. We utilized a commercial residual gas analyzer (RGA) to serve as an ion filter and detector (Extorr XT300M). The ions were detected by a channeltron in combination with an electron multiplier. The mass filter from the RGA has been used to guide the ions ejected from our ion trap towards the channeltron. The RGA allows filtering ions up to 300 amu. To make the RGA serve as a detector we removed the grid and filament at the RGA's entrance. Normally, these elements ionize the gas to be analyzed, but in our case, we wanted to detect already ionized atoms. When the grid and the filament were removed, the ions from our trap were able to enter freely and to be guided by the mass filter of the RGA.  We also added a switching mechanism for the RF field to RGA rods. In this way, we can switch off the filtering in RGA completely and filter charge states only with our Paul trap. The scheme in Fig.~\ref{fig:RGA_ion} illustrates the relative positions of the ion trap and detection system elements. 

\begin{figure}
    \centering
    \includegraphics[width=\linewidth]{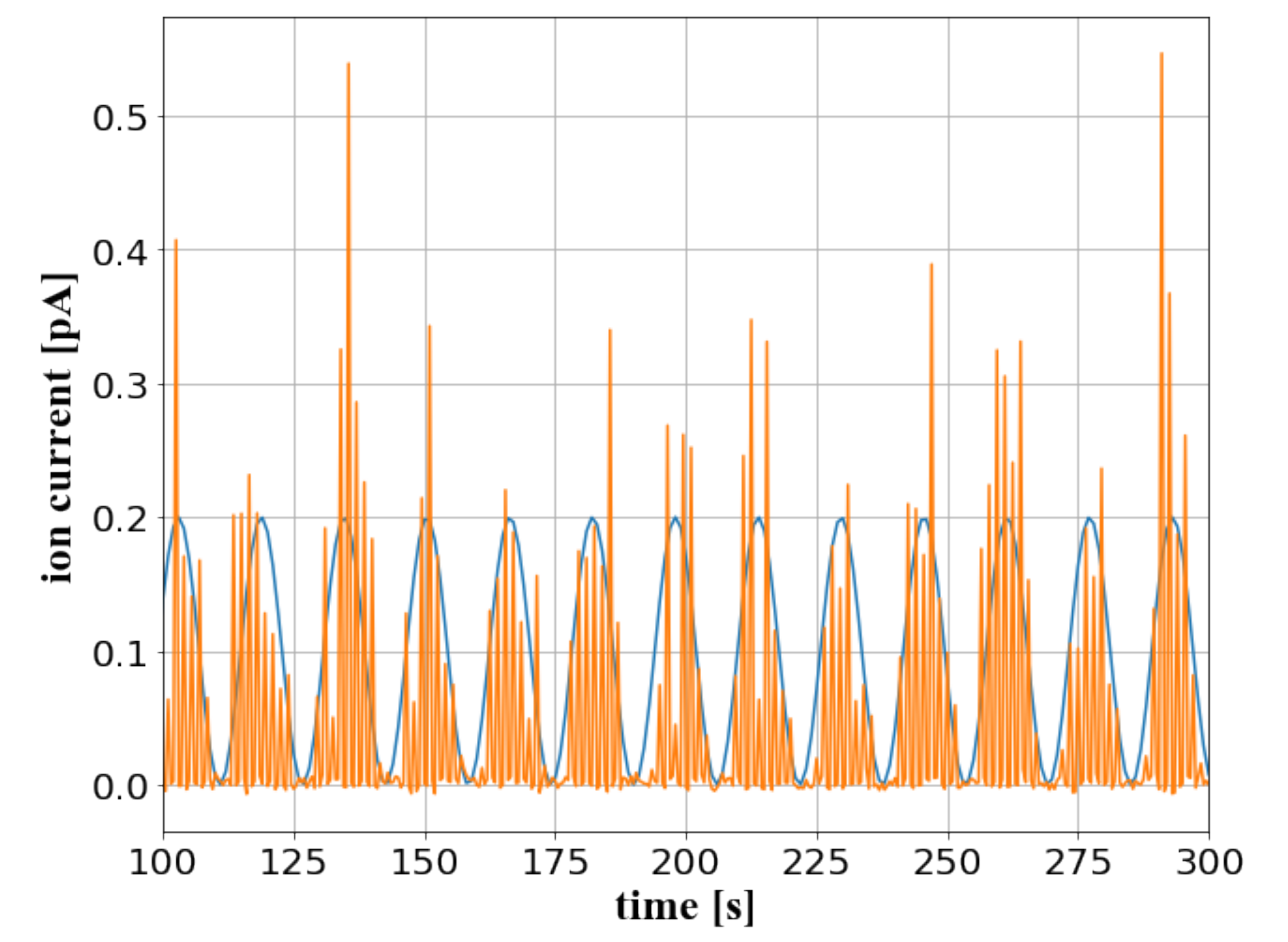}
    \caption{Example of the ion current detected in the channeltron as a function of time (green lines). The repetition rate of the ablation laser is 1~Hz, the ion trap driving frequency 6.24 MHz, ion signal data points were averaged over 500ms. The blue line is a fit with periodic signal $\mathrm{sin}^2(2\pi f+\phi)$ with frequency $2\pi f=198$~mHz corresponding to the beat frequency between ablation and trap frequency.}
    \label{fig:RF-phase}
\end{figure}

We observed that the number of ions guided by the quadrupole mass filter which reached the detector varied periodically. The first explanation of this behavior might be the phase difference of trap drive frequency and the repetition rate of the ablation laser, as the phases are not synchronized. In our case, the repetition rate of the pulsed ablation laser can be tuned between 1 -- 20Hz. An example of the ion current detected in the channeltron as a function of time (green lines) is shown in figure \ref{fig:RF-phase}. Another explanation might be that the flux of ions initially shortens the quadrupole rods to the ground and the following ions can be guided by the quadrupole field. A similar effect was observed in experiments investigating the RF trap loading with Ca ions \cite{hashimoto_trapping_2006}. The first of these potential explanations is more likely as we noticed the oscillation period depends on the trap frequency and repetition rate of the laser, but it might be also a combination of both. 

\section{Results}

In our experimental setup, we observed laser ablation from metallic targets of Yb and Th. The two ablation plumes could be visually easily distinguished - blue fluorescence from the plasma plume of Yb and white from Th. The ytterbium was much brighter to the eye and lasted longer, while the thorium ablation was much fainter and short-lived. It is consistent with the expected higher threshold for laser ablation of thorium. To develop a thorough understanding of these observations we present here a detailed quantitative analysis.

\subsection{Trapping and detection of Yb+, Th+ and Th2+ states}

\begin{figure}
    \centering
    \includegraphics[width=\linewidth]{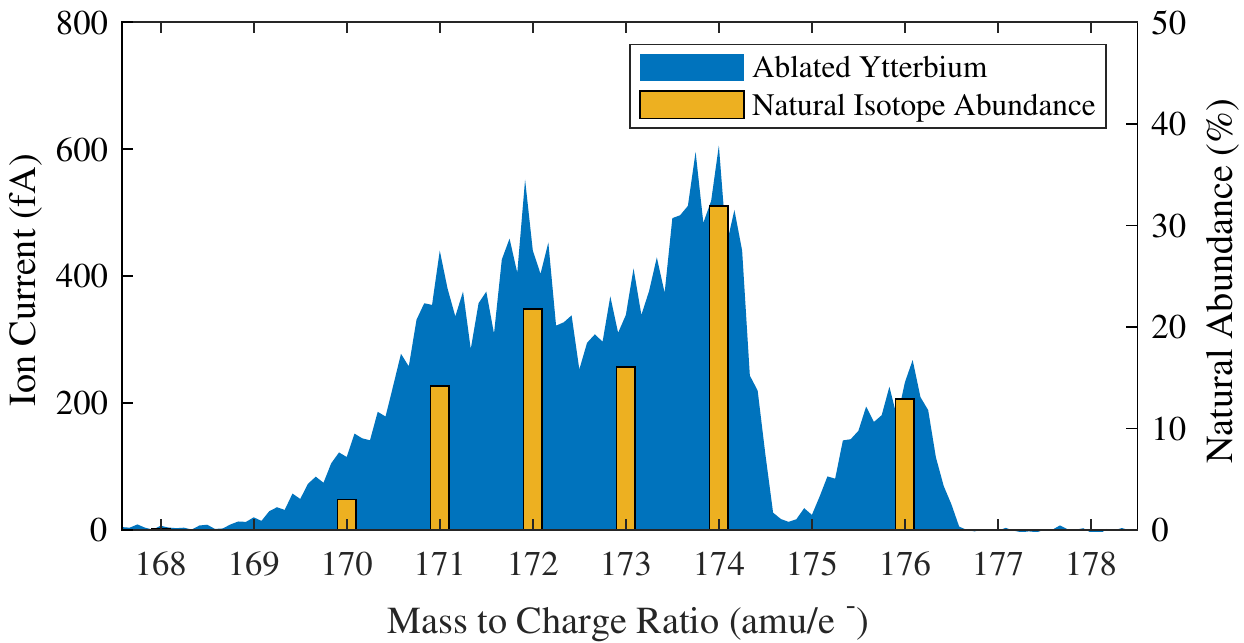}
    \caption{ Comparison of the Yb mass spectrum detected in the RGA with the natural isotope abundance of Yb.}
    \label{fig:Yb-isotopes}
\end{figure}

The laser ablation plume contains both neutral and ionized atoms, the ratio of the constituents depends on the energy density of the ablation laser. The first check of our ablation setup was to test it on a ytterbium target without the ion trap. The produced plasma plume was first analyzed by the unmodified RGA. The products of ablation were ionized by the RGA filament and were accelerated by the entrance grid voltages.  This way we reproduced the expected natural isotope abundance of ytterbium, but the detection, in this case, does not resolve neutral and ionized products of the ablation. The graph in figure \ref{fig:Yb-isotopes} shows the result of this measurement with Yb atoms ionized at the RGA entrance. 

\begin{figure}
\centering
    \includegraphics[width=\linewidth]{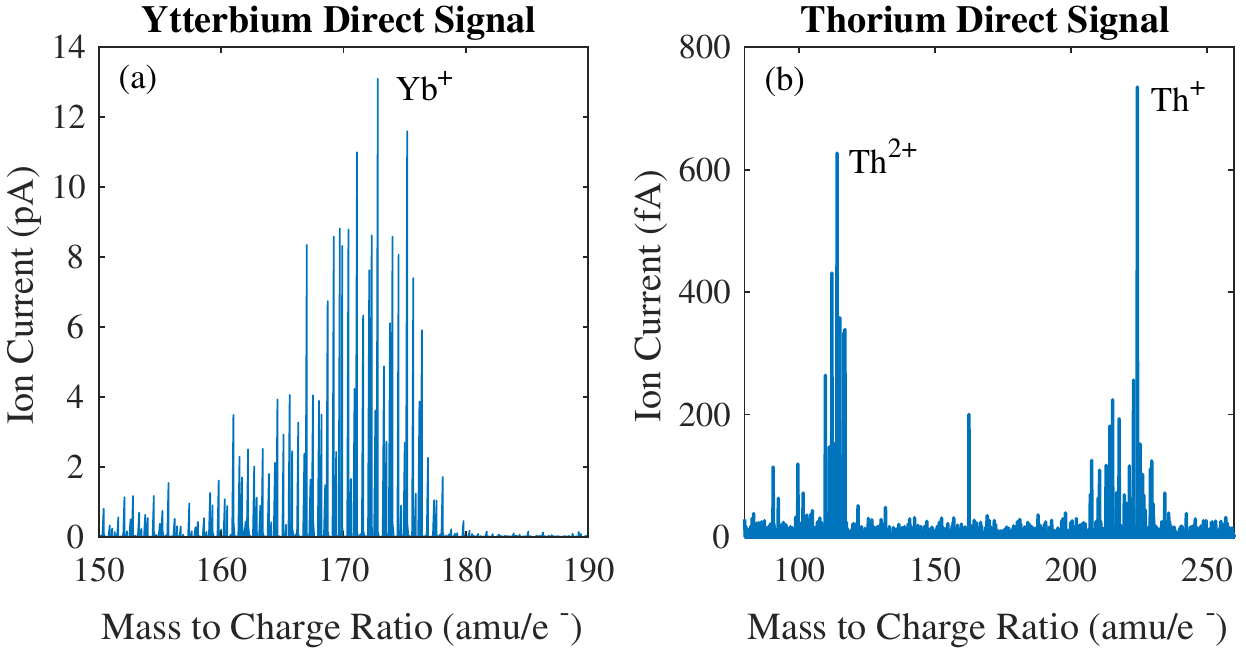}
    \caption{A signal confirming successful creation and RF-trap loading of (a)~Yb$^+$ and (b)~Th$^+$ and Th$^{2+}$ ions using our laser ablation method and detection scheme. The plots show ion current in the electron multiplier during the m/z scan of the RGA detector (horizontal axes).}
    \label{fig:Yb-Th-ions-from-trap}
\end{figure}

We also confirmed the signal from all isotopes of ytterbium without the high voltage and filament at the entrance of the RGA, while only measuring the ions produced by the laser ablation. We detected a clear signal from Yb$^{1+}$, Th$^{1+}$ and Th$^{2+}$ when the ion trap before the RGA transmitted particular charge states and the RGA scanned around mass to charge ratio during the laser ablation. These results are presented in figure \ref{fig:Yb-Th-ions-from-trap}. We have only confirmed a signal for singly and doubly ionized thorium.  Each point represented as a vertical line is an integration over 50 ms.

The first assumption about the threshold for the laser-produced plasma is the energy fluence required for vaporization from the metal surface. 
The threshold for thorium (approximated around $1~\mathrm{J/cm^2}$) is based on a purely thermal model of laser ablation \cite{chichkov_femtosecond_1996}.
Other models predict that the volume of the material ablated from the metal surface scales linearly with pulse energy \cite{leitz_metal_2011}.  We might expect that the fluence required for the ionization of metal atoms would be even higher; some energy from the pulse must be absorbed by the atoms to release electrons. To estimate the threshold for ionization, the energy density in the pulse must be compared with the ionization energy. We estimate our peak fluence to be only a fraction of the vaporization threshold, less than $0.7~\mathrm{J/cm^2}$. 
Regardless of the nominally sub-threshold fluence level, we observed the ablation of thorium targets and the creation of singly and doubly ionized atoms similarly to \cite{Zimmermann:2010pcc} where authors reached $1.1~\mathrm{J/cm^2}$ with the same kind of laser for ablation.
Successful trap loading $^{232}$Th$^{3+}$ has been reported in \cite{Borisyuk-2017} but with a more powerful laser source and fluence reaching $320~\mathrm{J/cm}^2$, calculated from their 50 mJ pulse energy and 100$~\mathrm{\mu m}$ laser spot size. A metallic thorium-232 target was used. Successful loading of $^{232}$Th$^{3+}$ via laser ablation has also been reported by Campbell et al \cite{PhysRevLett.102.233004} with a fluence of $60^{+460}_{-40}~\mathrm{J/cm}^2$, calculated from their 200$~\mathrm{\mu J}$ pulse energy and 15(10)$~\mathrm{\mu m}$ spot size. Large uncertainty in the fluence propagates from laser spot size uncertainty. Both groups used a thorium-232 metal target. Campbell et al \cite{Campbell2011,Campbell2011_2} later used a thorium nitrate sample and successfully loaded $^{229}$Th$^{3+}$ with a fluence of $16^{+48}_{-8}~\mathrm{J/cm^2}$ calculated from their 100$~\mathrm{\mu J}$ pulse energy and 20(10)$~\mathrm{\mu m}$ spot size.

\subsection{\label{sec:level1}Stability plots for thorium charge states}

\begin{figure}
\centering
\includegraphics[width=\linewidth]{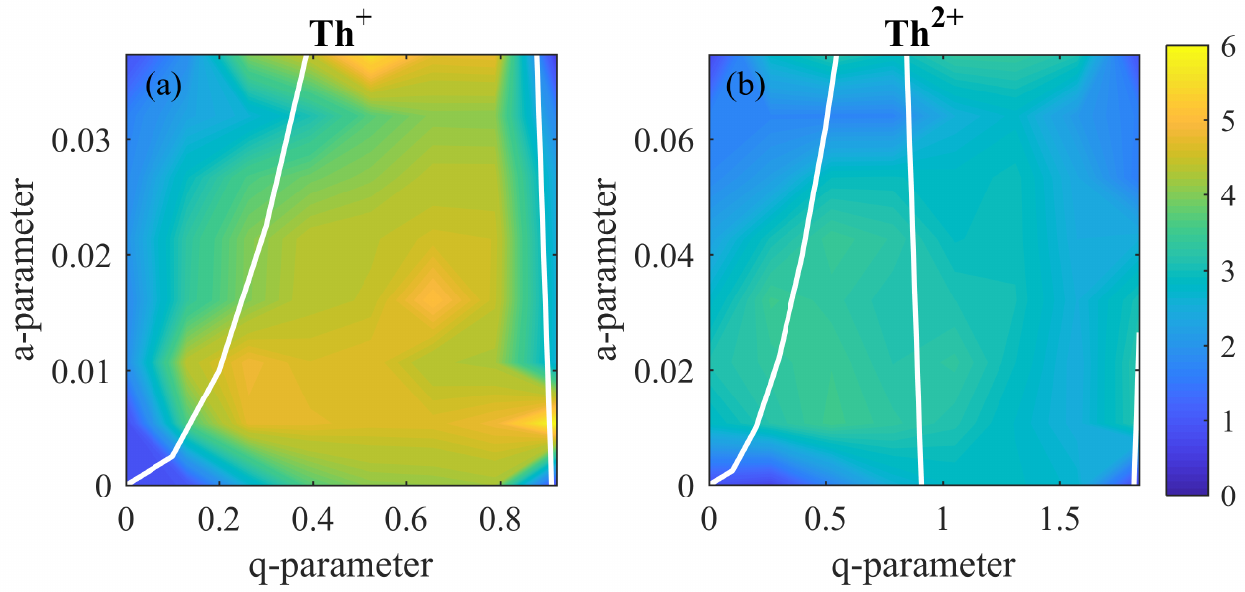}
    
\caption{The contour plots of the ion current as a function of the parameters $a$ and $q$ of the trap for Th$^{1+}$ and Th$^{2+}$ charge states. The horizontal axes represent the $q$~parameter proportional to the amplitude of the driving voltage of the ion trap at 6.2 MHz. The vertical axes represent $a$~parameter linked to the DC bias voltage applied to the trap electrodes. The color bar indicates the ion current at the detector (arbitrary log scale), a measure of the number of ions that traveled through the ion trap.  The predicted boundaries of stability regions from Mathieu equations are indicated by white lines.}  
    \label{fig:stability-plots}
\end{figure}

We scanned the RF voltage amplitude ($V_{RF}$) and DC bias voltage applied to the ion trap electrodes ($U_{DC}$) while the RGA mass filter was only transmitting to explore the parameter space of our trap. The radial motion of the ions on the quadruple trap is governed by the Mathieu equation \cite{RevModPhys.62.531} and we can extract two parameters called $a$ and $q$ that are proportional to $U_{DC}$ and $V_{RF}$ and other constant trap parameters. The expected scaling between charge states of thorium was confirmed. The trap parameters were scanned in 20 V steps for both RF and DC fields, while the detector parameters were fixed. For each data point, we measured the corresponding ion current.
The results are presented in figure \ref{fig:stability-plots}. The amplitude ($V_{RF}$) is inferred from the amplitude of the signal generator, accounting for the amplifier and independently measured losses in the helical resonator.

 The results show that, while there may be a systematic shift in the overall scale due to incomplete quantitative characterization of the trap, the stability regions do scale as expected when comparing the behavior of different m/z ratios.  Of particular importance is the scaling between Th$^{2+}$ and Th$^{1+}$ stability regions, which appear to follow the expected factor of two in the RF and DC voltages.  Despite a rigorous search, we were unable to detect any signal from Th$^{3+}$.

\subsection{\label{sec:level2}Laser ablation yield evolution}

\begin{figure}
    \centering
    \includegraphics[width=\linewidth]{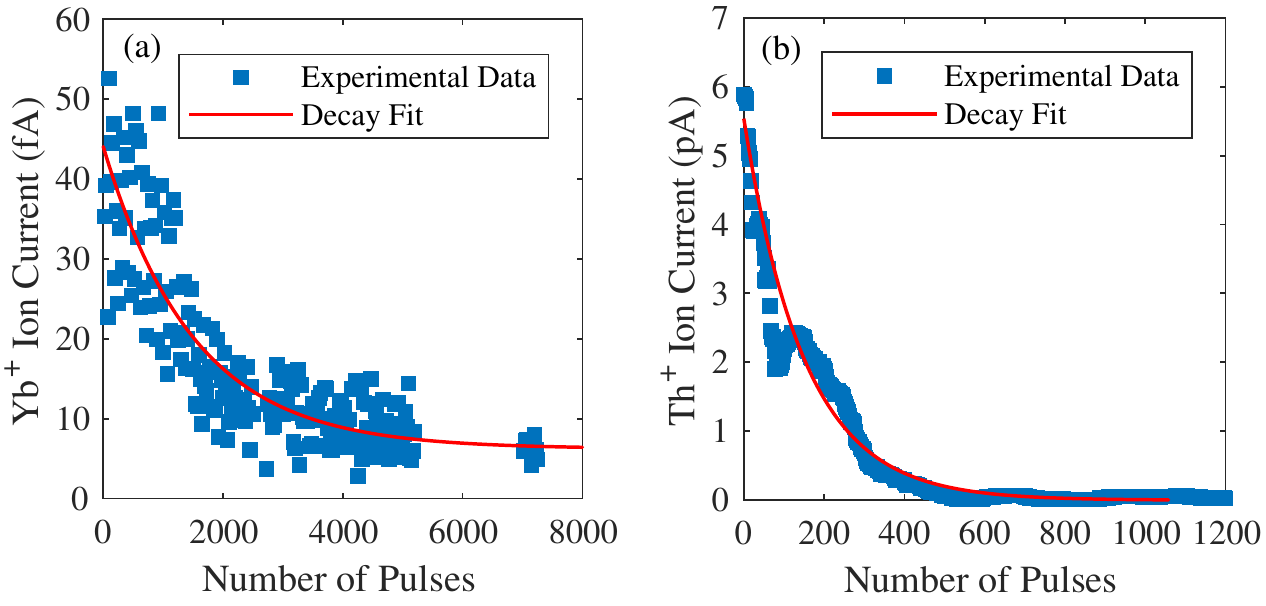}
    \caption{Time dependence of the ytterbium (a) and thorium (b) ions yield.  The laser is moved to an undamaged location on the ablation target, and a sequence of pulses is repeated on the same spot.}
    \label{fig:ablation-decay}
\end{figure}

The yield of the ablation changes over the number of pulses in the same spot on the target surface. The ablation decays are shown in Fig.~\ref{fig:ablation-decay} for the ytterbium (a) and thorium (b) targets. The ablation laser runs at 1~Hz rate with an integration time of 500 ms in our detector. The blue points represent the data after binning and averaging (bin size of 25 points for Yb and 5 points for Th), while the red line is the fitted curve of the exponential decay. The experimental runs with higher repetition rates and shorter exposures were noisier but showed similar behavior. 

The observation of a significant decrease in the ablation yield is contrary to previous studies which have suggested that the decay process is not significant and for solid metal targets ablation yield remains constant \cite{olmschenk_laser_2017}. Some results of ablation yield evolution for Sr, Yb and their oxides were also reported in \cite{zimmermann_laser_2012}, the authors recorded the yield evolution for Yb up to 500 pulses and did not observe any significant decay. In contrast to these reports, we have monitored laser ablation for much longer, up to a few thousand pulses. The significant drop in the yield amount was observed after 1500 pulses (reduction by $1/e$), which is consistent with previous studies; the authors could have missed a yield drop when observing only 500 pulses. However, the exponential decay for the thorium target was evident much more quickly ($1/e$ decay after ~150 pulses). This might seem surprising unless we consider that the ablation only comes from the oxide layer on the metal surface. That explains why it disappears so quickly, and the decay is in good agreement with similar observation for Yb and Sr oxides \cite{zimmermann_laser_2012}. The thermal diffusivity (which controls the heating characteristics) of metals is usually an order of magnitude higher than of oxides \cite{Doeswijk2004}, so the expected fluence threshold for oxides is also lower. That may explain the fact that what we ablate is not the thorium metal, but rather the layer of oxides on the surface. The thorium ions could be created during the plasma plume evolution afterward. 

\section{Conclusions}

We have suggested utilizing nuclear clocks based on the isomeric transition of the $^{229}$Th nucleus for chronometric geodesy and discussed experimental efforts toward its realization.
An ion trap for trapping and storing thorium ions was designed and built along with a vacuum enclosure and detection system. At the initial stage of research, we investigated methods of delivering thorium ions into the trap. We conducted a series of laser ablation experiments to determine a reliable and efficient way to deliver ions from metal targets. We showed that even a simple and fairly weak ablation laser can lead to thorium ionization lasting long enough to be effectively injected into an RF trap. Our results are in good agreement with previously reported trap loading with thorium, but the extended studies of yield from laser ablation suggest that only the oxide layers contribute to the Th ion production by UV nitrogen laser pulses. The findings are important for optimal loading of miniaturized thorium traps,  which are expected to be the crucial component of future nuclear clocks.  

We studied electroplating for thorium target preparation. $^{232}$Th was used in this study as a predecessor for $^{229}$Th, for reasons of lower radioactivity. Our findings can be directly transferred to the $^{229}$Th isotope. Our laser setup cannot produce the fluence required for the ablation of triply ionized thorium, thus more powerful laser must be employed for this task or the technique presented combined with other means of ionization like photo-ionization or electron bombardment. We also plan to add time gating of the trap for isotope selectivity. Once we produce and trap Th$^{3+}$, we can perform laser cooling working towards the nuclear clock transition spectroscopy.

\section*{Funding}
This work was funded by the Commonwealth Scientific Industrial Research Organisation (CSIRO). Mirko Lobino was supported by the Australian Research Council Future Fellowship FT180100055. Erik Streed was supported by Australian Research Council Future Fellowship FT130100472 and Linkage Project LP180100096.

\section*{Acknowledgments}
Laboratory space and support services were provided by Griffith University.  We thank Dane Laban for his advice in the development of the thorium deposition apparatus.

\section*{Disclosures}
The authors declare no conflicts of interest.


\bibliography{books,laser-ablation,geopotential-clocks,ablation-2}

\begin{thebibliography}{10}

\bibitem{vermeer_chronometric_1983}
M.~Vermeer, {\em Chronometric {Levelling}}.
\newblock Reports of the {Finnish} {Geodetic} {Institute}, Geodeettinen Laitos,
  Geodetiska Institutet, 1983.

\bibitem{0034-4885-81-6-064401}
T.~E. Mehlst{\"a}ubler, G.~Grosche, C.~Lisdat, P.~O. Schmidt, and H.~Denker,
  ``Atomic clocks for geodesy,'' {\em Reports on Progress in Physics}, vol.~81,
  no.~6, p.~064401, 2018.

\bibitem{takano_geopotential_2016}
T.~Takano, M.~Takamoto, I.~Ushijima, N.~Ohmae, T.~Akatsuka, A.~Yamaguchi,
  Y.~Kuroishi, H.~Munekane, B.~Miyahara, and H.~Katori, ``Geopotential
  measurements with synchronously linked optical lattice clocks,'' {\em Nature
  Photonics}, vol.~10, p.~662, Oct. 2016.

\bibitem{lisdat_clock_2016}
C.~Lisdat, G.~Grosche, N.~Quintin, C.~Shi, S.~M.~F. Raupach, C.~Grebing,
  D.~Nicolodi, F.~Stefani, A.~Al-Masoudi, S.~D{\"o}rscher, S.~H{\"a}fner, J.-L.
  Robyr, N.~Chiodo, S.~Bilicki, E.~Bookjans, A.~Koczwara, S.~Koke, A.~Kuhl,
  F.~Wiotte, F.~Meynadier, E.~Camisard, M.~Abgrall, M.~Lours, T.~Legero,
  H.~Schnatz, U.~Sterr, H.~Denker, C.~Chardonnet, Y.~L. Coq, G.~Santarelli,
  A.~Amy-Klein, R.~L. Targat, J.~Lodewyck, O.~Lopez, and P.-E. Pottie, ``A
  clock network for geodesy and fundamental science,'' {\em Nature
  Communications}, vol.~7, p.~12443, Aug. 2016.

\bibitem{chou_optical_2010}
C.~W. Chou, D.~B. Hume, T.~Rosenband, and D.~J. Wineland, ``Optical {Clocks}
  and {Relativity},'' {\em Science}, vol.~329, pp.~1630--1633, Sept. 2010.

\bibitem{PhysRevLett.118.221102}
P.~Delva, J.~Lodewyck, S.~Bilicki, E.~Bookjans, G.~Vallet, R.~Le~Targat, P.-E.
  Pottie, C.~Guerlin, F.~Meynadier, C.~Le~Poncin-Lafitte, O.~Lopez,
  A.~Amy-Klein, W.-K. Lee, N.~Quintin, C.~Lisdat, A.~Al-Masoudi, S.~D\"orscher,
  C.~Grebing, G.~Grosche, A.~Kuhl, S.~Raupach, U.~Sterr, I.~R. Hill, R.~Hobson,
  W.~Bowden, J.~Kronj\"ager, G.~Marra, A.~Rolland, F.~N. Baynes, H.~S.
  Margolis, and P.~Gill, ``Test of special relativity using a fiber network of
  optical clocks,'' {\em Phys. Rev. Lett.}, vol.~118, p.~221102, Jun 2017.

\bibitem{Grotti2018}
J.~Grotti, S.~Koller, S.~Vogt, S.~H{\"a}fner, U.~Sterr, C.~Lisdat, H.~Denker,
  C.~Voigt, L.~Timmen, A.~Rolland, F.~N. Baynes, H.~S. Margolis, M.~Zampaolo,
  P.~Thoumany, M.~Pizzocaro, B.~Rauf, F.~Bregolin, A.~Tampellini, P.~Barbieri,
  M.~Zucco, G.~A. Costanzo, C.~Clivati, F.~Levi, and D.~Calonico, ``Geodesy and
  metrology with a transportable optical clock,'' {\em Nature Physics},
  vol.~14, no.~5, pp.~437--441, 2018.

\bibitem{McGrew2018}
W.~F. McGrew, X.~Zhang, R.~J. Fasano, S.~A. Sch{\"a}ffer, K.~Beloy,
  D.~Nicolodi, R.~C. Brown, N.~Hinkley, G.~Milani, M.~Schioppo, T.~H. Yoon, and
  A.~D. Ludlow, ``Atomic clock performance enabling geodesy below the
  centimetre level,'' {\em Nature}, vol.~564, pp.~87--90, Dec 2018.

\bibitem{Predehl441}
K.~Predehl, G.~Grosche, S.~M.~F. Raupach, S.~Droste, O.~Terra, J.~Alnis,
  T.~Legero, T.~W. H{\"a}nsch, T.~Udem, R.~Holzwarth, and H.~Schnatz, ``A
  920-kilometer optical fiber link for frequency metrology at the 19th decimal
  place,'' {\em Science}, vol.~336, no.~6080, pp.~441--444, 2012.

\bibitem{Lopez:12}
O.~Lopez, A.~Haboucha, B.~Chanteau, C.~Chardonnet, A.~Amy-Klein, and
  G.~Santarelli, ``Ultra-stable long distance optical frequency distribution
  using the internet fiber network,'' {\em Opt. Express}, vol.~20,
  pp.~23518--23526, Oct 2012.

\bibitem{Brewer_2019}
S.~M. Brewer, J.-S. Chen, A.~M. Hankin, E.~R. Clements, C.~W. Chou, D.~J.
  Wineland, D.~B. Hume, and D.~R. Leibrandt, ``$^{27}${A}l$^{+}$ quantum-logic
  clock with a systematic uncertainty below ${10}^{\ensuremath{-}18}$,'' {\em
  Phys. Rev. Lett.}, vol.~123, p.~033201, Jul 2019.

\bibitem{koller_transportable_2017}
S.~Koller, J.~Grotti, S.~Vogt, A.~Al-Masoudi, S.~D{\"o}rscher, S.~H{\"a}fner,
  U.~Sterr, and C.~Lisdat, ``Transportable optical lattice clock with
  $7\ifmmode\times\else\texttimes\fi{}{10}^{\ensuremath{-}17}$ uncertainty,''
  {\em Physical Review Letters}, vol.~118, p.~073601, Feb. 2017.

\bibitem{takamoto_test_2020}
M.~Takamoto, I.~Ushijima, N.~Ohmae, T.~Yahagi, K.~Kokado, H.~Shinkai, and
  H.~Katori, ``Test of general relativity by a pair of transportable optical
  lattice clocks,'' {\em Nature Photonics}, pp.~1--5, Apr. 2020.

\bibitem{Beck_energy_2007}
B.~R. Beck, J.~A. Becker, P.~Beiersdorfer, G.~V. Brown, K.~J. Moody, J.~B.
  Wilhelmy, F.~S. Porter, C.~A. Kilbourne, and R.~L. Kelley, ``Energy splitting
  of the ground-state doublet in the nucleus $^{229}\mathrm{Th}$,'' {\em Phys.
  Rev. Lett.}, vol.~98, p.~142501, Apr 2007.

\bibitem{Beck_imroved_2009}
B.~R. Beck, C.~Wu, P.~Beiersdorfer, G.~V. Brown, J.~A. Becker, K.~J. Moody,
  J.~B. Wilhelmy, F.~S. Porter, C.~A. Kilbourne, and R.~L. Kelley, ``Improved
  value for the energy splitting of the ground-state doublet in the nucleus
  {$^{229\mathrm{m}}$Th},'' No.~LLNL-PROC-415170, 2009.

\bibitem{Seiferle2019}
B.~Seiferle, L.~von~der Wense, P.~V. Bilous, I.~Amersdorffer, C.~Lemell,
  F.~Libisch, S.~Stellmer, T.~Schumm, C.~E. D{\"u}llmann, A.~P{\'a}lffy, and
  P.~G. Thirolf, ``Energy of the {$^{229}$Th} nuclear clock transition,'' {\em
  Nature}, vol.~573, pp.~243--246, Sep 2019.

\bibitem{Yamaguchi_2019}
A.~Yamaguchi, H.~Muramatsu, T.~Hayashi, N.~Yuasa, K.~Nakamura, M.~Takimoto,
  H.~Haba, K.~Konashi, M.~Watanabe, H.~Kikunaga, K.~Maehata, N.~Y. Yamasaki,
  and K.~Mitsuda, ``Energy of the $^{229}\mathrm{Th}$ nuclear clock isomer
  determined by absolute $\ensuremath{\gamma}$-ray energy difference,'' {\em
  Phys. Rev. Lett.}, vol.~123, p.~222501, Nov 2019.

\bibitem{sikorsky2020measurement}
{Tomas Sikorsky}, {Jeschua Geist}, {Daniel Hengstler}, {Sebastian Kempf},
  {Loredana Gastaldo}, {Christian Enss}, {Christoph Mokry}, {J{\"{o}}rg Runke},
  {Christoph E. D{\"{u}}llmann}, {Peter Wobrauschek}, {Kjeld Beeks}, {Veronika
  Rosecker}, {Johannes H. Sterba}, {Georgy Kazakov}, {Thorsten Schumm}, and
  {Andreas Fleischmann}, ``{Measurement of the $^{229}$Th isomer energy with a
  magnetic micro-calorimeter}.'' \url{https://arxiv.org/abs/2005.13340}, may
  2020.

\bibitem{KROGER1976}
L.~Kroger and C.~Reich, ``Features of the low-energy level scheme of
  {$^{229}$Th} as observed in the $\alpha$-decay of {233U},'' {\em Nuclear
  Physics A}, vol.~259, no.~1, pp.~29 -- 60, 1976.

\bibitem{Rellergert2010}
W.~G. Rellergert, D.~Demille, R.~R. Greco, M.~P. Hehlen, J.~R. Torgerson, and
  E.~R. Hudson, ``{Constraining the evolution of the fundamental constants with
  a solid-state optical frequency reference based on the $^{229}$Th nucleus},''
  {\em Physical Review Letters}, vol.~104, p.~200802, may 2010.

\bibitem{Porsev2010}
S.~G. Porsev, V.~V. Flambaum, E.~Peik, and C.~Tamm, ``{Excitation of the
  isomeric $^{229\mathrm{m}}$Th nuclear state via an electronic bridge process
  in $^{229}$Th$^+$},'' {\em Physical Review Letters}, vol.~105, p.~182501, oct
  2010.

\bibitem{VonDerWense2017}
L.~{Von Der Wense}, B.~Seiferle, S.~Stellmer, J.~Weitenberg, G.~Kazakov,
  A.~P{\'{a}}lffy, and P.~G. Thirolf, ``{A Laser Excitation Scheme for
  $^{229\mathrm{m}}${Th}},'' {\em Physical Review Letters}, vol.~119,
  p.~132503, sep 2017.

\bibitem{Verlinde2019}
M.~Verlinde, S.~Kraemer, J.~Moens, K.~Chrysalidis, J.~G. Correia, S.~Cottenier,
  H.~{De Witte}, D.~V. Fedorov, V.~N. Fedosseev, R.~Ferrer, L.~M. Fraile,
  S.~Geldhof, C.~A. Granados, M.~Laatiaoui, T.~A. Lima, P.~C. Lin, V.~Manea,
  B.~A. Marsh, I.~Moore, L.~M. Pereira, S.~Raeder, P.~{Van Den Bergh}, P.~{Van
  Duppen}, A.~Vantomme, E.~Verstraelen, U.~Wahl, and S.~G. Wilkins,
  ``{Alternative approach to populate and study the Th 229 nuclear clock
  isomer},'' {\em Physical Review C}, vol.~100, p.~24315, aug 2019.

\bibitem{PhysRevC.100.044306}
P.~V. Borisyuk, N.~N. Kolachevsky, A.~V. Taichenachev, E.~V. Tkalya, I.~Y.
  Tolstikhina, and V.~I. Yudin, ``Excitation of the low-energy
  $^{229\mathrm{m}}\mathrm{Th}$ isomer in the electron bridge process via the
  continuum,'' {\em Phys. Rev. C}, vol.~100, p.~044306, Oct 2019.

\bibitem{Masuda2019}
T.~Masuda, A.~Yoshimi, A.~Fujieda, H.~Fujimoto, H.~Haba, H.~Hara, T.~Hiraki,
  H.~Kaino, Y.~Kasamatsu, S.~Kitao, K.~Konashi, Y.~Miyamoto, K.~Okai, S.~Okubo,
  N.~Sasao, M.~Seto, T.~Schumm, Y.~Shigekawa, K.~Suzuki, S.~Stellmer,
  K.~Tamasaku, S.~Uetake, M.~Watanabe, T.~Watanabe, Y.~Yasuda, A.~Yamaguchi,
  Y.~Yoda, T.~Yokokita, M.~Yoshimura, and K.~Yoshimura, ``{X-ray pumping of the
  229Th nuclear clock isomer},'' {\em Nature}, vol.~573, pp.~238--242, sep
  2019.

\bibitem{Bilous2020}
P.~V. Bilous, H.~Bekker, J.~C. Berengut, B.~Seiferle, L.~{Von Der Wense}, P.~G.
  Thirolf, T.~Pfeifer, J.~R. L{\'{o}}pez-Urrutia, and A.~P{\'{a}}lffy,
  ``{Electronic Bridge Excitation in Highly Charged Th 229 Ions},'' {\em
  Physical Review Letters}, vol.~124, p.~192502, may 2020.

\bibitem{VonderWense2020}
L.~von~der Wense and C.~Zhang, ``{Concepts for direct frequency-comb
  spectroscopy of $^{229\mathrm{m}}$Th and an internal-conversion-based
  solid-state nuclear clock},'' {\em The European Physical Journal D}, vol.~74,
  no.~7, p.~146, 2020.

\bibitem{Porsev2010_2}
S.~G. Porsev and V.~V. Flambaum, ``{Effect of atomic electrons on the 7.6-eV
  nuclear transition in Th2293 +},'' {\em Physical Review A - Atomic,
  Molecular, and Optical Physics}, vol.~81, p.~032504, mar 2010.

\bibitem{Bilous2018}
P.~V. Bilous, N.~Minkov, and A.~P{\'{a}}lffy, ``{Electric quadrupole channel of
  the 7.8 eV Th 229 transition},'' {\em Physical Review C}, vol.~97, p.~044320,
  apr 2018.

\bibitem{Peik_2003}
E.~Peik and C.~Tamm, ``Nuclear laser spectroscopy of the 3.5 {eV} transition in
  $\mathrm{Th}$-229,'' {\em Europhysics Letters ({EPL})}, vol.~61,
  pp.~181--186, jan 2003.

\bibitem{campbell_single-ion_2012}
C.~J. Campbell, A.~G. Radnaev, A.~Kuzmich, V.~A. Dzuba, V.~V. Flambaum, and
  A.~Derevianko, ``Single-{Ion} {Nuclear} {Clock} for {Metrology} at the 19th
  {Decimal} {Place},'' {\em Physical Review Letters}, vol.~108, p.~120802, Mar.
  2012.

\bibitem{Delehaye-2018}
M.~Delehaye and C.~Lacro{\^u}te, ``Single-ion, transportable optical atomic
  clocks,'' {\em Journal of Modern Optics}, vol.~65, no.~5-6, pp.~622--639,
  2018.

\bibitem{PEIK2015516}
E.~Peik and M.~Okhapkin, ``Nuclear clocks based on resonant excitation of
  $\gamma$-transitions,'' {\em Comptes Rendus Physique}, vol.~16, no.~5,
  pp.~516 -- 523, 2015.

\bibitem{Flambaum_enhenced_2006}
V.~V. Flambaum, ``Enhanced effect of temporal variation of the fine structure
  constant and the strong interaction in $^{229}\mathrm{Th}$,'' {\em Phys. Rev.
  Lett.}, vol.~97, p.~092502, Aug 2006.

\bibitem{Berengut_Flambaum_2010}
J.~C. Berengut and V.~V. Flambaum, ``Testing time-variation of fundamental
  constants using a $^{229}${Th} nuclear clock,'' {\em Nuclear Physics News},
  vol.~20, no.~3, pp.~19--22, 2010.

\bibitem{Thirolf_2019}
P.~G. Thirolf, B.~Seiferle, and L.~von~der Wense, ``Improving our knowledge on
  the $^{229\mathrm{m}}${T}horium isomer: Toward a test bench for time
  variations of fundamental constants,'' {\em Annalen der Physik}, vol.~531,
  no.~5, p.~1800381, 2019.

\bibitem{PhysRevLett.102.233004}
C.~J. Campbell, A.~V. Steele, L.~R. Churchill, M.~V. DePalatis, D.~E. Naylor,
  D.~N. Matsukevich, A.~Kuzmich, and M.~S. Chapman, ``Multiply charged thorium
  crystals for nuclear laser spectroscopy,'' {\em Phys. Rev. Lett.}, vol.~102,
  p.~233004, Jun 2009.

\bibitem{Campbell2011}
C.~J. Campbell, A.~G. Radnaev, and A.~Kuzmich, ``{Wigner crystals of Th229 for
  optical excitation of the nuclear isomer},'' {\em Physical Review Letters},
  vol.~106, p.~223001, jun 2011.

\bibitem{1807.05975}
K.~Groot-Berning, F.~Stopp, G.~Jacob, D.~Budker, R.~Haas, D.~Renisch, J.~Runke,
  P.~Th\"orle-Pospiech, C.~E. D\"ullmann, and F.~Schmidt-Kaler, ``Trapping and
  sympathetic cooling of single thorium ions for spectroscopy,'' {\em Phys.
  Rev. A}, vol.~99, p.~023420, Feb 2019.

\bibitem{Borisyuk-2017}
P.~V. Borisyuk, S.~P. Derevyashkin, K.~Y. Khabarova, N.~N. Kolachevsky, Y.~Y.
  Lebedinsky, S.~S. Poteshin, A.~A. Sysoev, E.~V. Tkalya, D.~O. Tregubov, V.~I.
  Troyan, O.~S. Vasiliev, V.~P. Yakovlev, and V.~I. Yudin, ``Loading of mass
  spectrometry ion trap with {Th} ions by laser ablation for nuclear frequency
  standard application,'' {\em European Journal of Mass Spectrometry}, vol.~23,
  no.~4, pp.~146--151, 2017.

\bibitem{Borisyuk-2017-2}
P.~V. Borisyuk, S.~P. Derevyashkin, K.~Y. Khabarova, N.~N. Kolachevsky, Y.~Y.
  Lebedinsky, S.~S. Poteshin, A.~A. Sysoev, E.~V. Tkalya, D.~O. Tregubov, V.~I.
  Troyan, O.~S. Vasiliev, V.~P. Yakovlev, and V.~I. Yudin, ``Mass selective
  laser cooling of {$^{229}$Th$^{3+}$} in a multisectional linear {Paul} trap
  loaded with a mixture of thorium isotopes,'' {\em European Journal of Mass
  Spectrometry}, vol.~23, no.~4, pp.~136--139, 2017.

\bibitem{PhysRevA.88.012512}
O.~A. Herrera-Sancho, N.~Nemitz, M.~V. Okhapkin, and E.~Peik, ``Energy levels
  of {Th}${}^{+}$ between 7.3 and 8.3 ev,'' {\em Phys. Rev. A}, vol.~88,
  p.~012512, Jul 2013.

\bibitem{PhysRevA.85.033402}
O.~A. Herrera-Sancho, M.~V. Okhapkin, K.~Zimmermann, C.~Tamm, E.~Peik, A.~V.
  Taichenachev, V.~I. Yudin, and P.~G{\l}owacki, ``Two-photon laser excitation
  of trapped ${}^{232}${Th}${}^{+}$ ions via the 402-nm resonance line,'' {\em
  Phys. Rev. A}, vol.~85, p.~033402, Mar 2012.

\bibitem{thielking_laser_2018}
J.~Thielking, M.~V. Okhapkin, P.~G{\l}owacki, D.~M. Meier, L.~v.~d. Wense,
  B.~Seiferle, C.~E. D{\"u}llmann, P.~G. Thirolf, and E.~Peik, ``Laser
  spectroscopic characterization of the nuclear-clock isomer
  $^{229\mathrm{m}}${Th},'' {\em Nature}, vol.~556, pp.~321--325, Apr. 2018.

\bibitem{Blumseaao4453}
V.~Blums, M.~Piotrowski, M.~I. Hussain, B.~G. Norton, S.~C. Connell,
  S.~Gensemer, M.~Lobino, and E.~W. Streed, ``A single-atom 3d sub-attonewton
  force sensor,'' {\em Science Advances}, vol.~4, no.~3, 2018.

\bibitem{Streed-unfolding}
E.~Streed, ``Unfolding large biomolecules,'' {\em Proceedings of the 2012
  Australian Institute of Physics Congress}, 11 2012.

\bibitem{willmott_pulsed_2000}
P.~R. Willmott and J.~R. Huber, ``Pulsed laser vaporization and deposition,''
  {\em Reviews of Modern Physics}, vol.~72, pp.~315--328, Jan. 2000.

\bibitem{knight_storage_1981}
R.~D. Knight, ``Storage of ions from laser produced plasmas,'' {\em Applied
  Physics Letters}, vol.~38, pp.~221--223, Feb. 1981.

\bibitem{olmschenk_laser_2017}
S.~Olmschenk and P.~Becker, ``Laser ablation production of {Ba}, {Ca}, {Dy},
  {Er}, {La}, {Lu}, and {Yb} ions,'' {\em Applied Physics B}, vol.~123, p.~99,
  Apr. 2017.

\bibitem{hendricks_all-optical_2007}
R.~J. Hendricks, D.~M. Grant, P.~F. Herskind, A.~Dantan, and M.~Drewsen, ``An
  all-optical ion-loading technique for scalable microtrap architectures,''
  {\em Applied Physics B}, vol.~88, pp.~507--513, Sept. 2007.

\bibitem{cao_compact_2017}
J.~Cao, P.~Zhang, J.~Shang, K.~Cui, J.~Yuan, S.~Chao, S.~Wang, H.~Shu, and
  X.~Huang, ``A compact, transportable single-ion optical clock with $7.8
  \times 10^{-17}$ systematic uncertainty,'' {\em Applied Physics B}, vol.~123,
  p.~112, Apr. 2017.

\bibitem{Vrijsen:19}
G.~Vrijsen, Y.~Aikyo, R.~F. Spivey, I.~V. Inlek, and J.~Kim, ``Efficient
  isotope-selective pulsed laser ablation loading of {174Yb+} ions in a surface
  electrode trap,'' {\em Opt. Express}, vol.~27, pp.~33907--33914, Nov 2019.

\bibitem{sheridan_all-optical_2011}
K.~Sheridan, W.~Lange, and M.~Keller, ``All-optical ion generation for ion trap
  loading,'' {\em Applied Physics B}, vol.~104, p.~755, Sept. 2011.

\bibitem{leibrandt_laser_2007}
D.~R. Leibrandt, R.~J. Clark, J.~Labaziewicz, P.~Antohi, W.~Bakr, K.~R. Brown,
  and I.~L. Chuang, ``Laser ablation loading of a surface-electrode ion trap,''
  {\em Physical Review A}, vol.~76, p.~055403, Nov. 2007.

\bibitem{zimmermann_laser_2012}
K.~Zimmermann, M.~V. Okhapkin, O.~A. Herrera-Sancho, and E.~Peik, ``Laser
  ablation loading of a radiofrequency ion trap,'' {\em Applied Physics B},
  vol.~107, pp.~883--889, June 2012.

\bibitem{troyan_generation_2013}
V.~I. Troyan, P.~V. Borisyuk, R.~R. Khalitov, A.~V. Krasavin, Y.~Y.
  Lebedinskii, V.~G. Palchikov, S.~S. Poteshin, A.~A. Sysoev, and V.~P.
  Yakovlev, ``Generation of thorium ions by laser ablation and inductively
  coupled plasma techniques for optical nuclear spectroscopy,'' {\em Laser
  Physics Letters}, vol.~10, no.~10, p.~105301, 2013.

\bibitem{RevModPhys.62.531}
W.~Paul, ``Electromagnetic traps for charged and neutral particles,'' {\em Rev.
  Mod. Phys.}, vol.~62, pp.~531--540, Jul 1990.

\bibitem{Kielpinski:06}
D.~Kielpinski, M.~Cetina, J.~A. Cox, and F.~X. K\"{a}rtner, ``Laser cooling of
  trapped ytterbium ions with an ultraviolet diode laser,'' {\em Opt. Lett.},
  vol.~31, pp.~757--759, Mar 2006.

\bibitem{Streed-2008}
E.~W. Streed, T.~J. Weinhold, and D.~Kielpinski, ``Frequency stabilization of
  an ultraviolet laser to ions in a discharge,'' {\em Applied Physics Letters},
  vol.~93, no.~7, p.~071103, 2008.

\bibitem{Dumitru2013}
O.~A. Dumitru, R.~C. Begy, D.~C. Nita, L.~D. Bobos, and C.~Cosma, ``Uranium
  electrodeposition for alpha spectrometric source preparation,'' {\em Journal
  of Radioanalytical and Nuclear Chemistry}, vol.~298, pp.~1335--1339, Nov
  2013.

\bibitem{hashimoto_trapping_2006}
Y.~Hashimoto, L.~Matsuoka, H.~Osaki, Y.~Fukushima, and S.~Hasegawa, ``Trapping
  {Laser} {Ablated} {Ca}+ {Ions} in {Linear} {Paul} {Trap},'' {\em Japanese
  Journal of Applied Physics}, vol.~45, p.~7108, Sept. 2006.

\bibitem{chichkov_femtosecond_1996}
B.~N. Chichkov, C.~Momma, S.~Nolte, F.~von Alvensleben, and A.~T{\"u}nnermann,
  ``Femtosecond, picosecond and nanosecond laser ablation of solids,'' {\em
  Applied Physics A}, vol.~63, pp.~109--115, Aug. 1996.

\bibitem{leitz_metal_2011}
K.-H. Leitz, B.~Redlingshofer, Y.~Reg, A.~Otto, and M.~Schmidt, ``Metal
  {Ablation} with {Short} and {Ultrashort} {Laser} {Pulses},'' {\em Physics
  Procedia}, vol.~12, pp.~230--238, Jan. 2011.

\bibitem{Zimmermann:2010pcc}
K.~Zimmermann, {\em {Experiments towards optical nuclear spectroscopy with
  Thorium-229}}.
\newblock PhD thesis, Leibniz U., Hannover, 2010.

\bibitem{Campbell2011_2}
C.~J. Campbell, {\em {Trapping, laser cooling, and spectroscopy of Thorium
  IV}}.
\newblock PhD thesis, Georgia Institute of Technology, jun 2011.

\bibitem{Doeswijk2004}
L.~M. Doeswijk, G.~Rijnders, and D.~H.~A. Blank, ``Pulsed laser deposition:
  metal versus oxide ablation,'' {\em Applied Physics A}, vol.~78,
  pp.~263--268, Feb 2004.

\end{thebibliography}

\end{document}